\documentstyle[12pt]{article}
\newcommand{\be}{\begin{equation}}
\newcommand{\ee}{\end{equation}}
\newcommand{\rref}[1]{(\ref{#1})}
\newcommand{\N}{{\cal N}}

\begin{document}

\begin{flushright}
ULB--TH--97/01 \\
hep-th/9701042 \\
January 1997\\
\end{flushright}

\vspace{.8cm}

\begin{center}
{\Huge Intersection Rules for $p$-Branes}
\vspace{1.5cm}

{\large R.~Argurio}${}^*$\footnote{
Aspirant F.N.R.S. (Belgium). E-mail:
rargurio@ulb.ac.be}{\large , F.~Englert}${}^{*\dagger}$\footnote{
E-mail: fenglert@ulb.ac.be}
{\large and L.~Houart}${}^*$\footnote{
Charg\'e de Recherches F.N.R.S. (Belgium). E-mail: lhouart@ulb.ac.be}
\addtocounter{footnote}{-3}\\
\vspace{.4cm}
${}^*${\it Service de Physique Th\'eorique}\\
{\it Universit\'e Libre de Bruxelles, Campus Plaine, C.P.225}\\
{\it Boulevard du Triomphe, B-1050 Bruxelles, Belgium}\\
\vspace{.4cm}
${}^\dagger${\it Raymond and Beverly Sackler Faculty of Exact Sciences}\\
{\it School of Physics and Astronomy}\\
{\it Tel Aviv University, Ramat Aviv, 69978, Israel}
\end{center}

\vspace{1.5cm}

\begin{abstract}

We present a general rule determining how extremal branes can interesect
in a configuration with zero binding energy. The rule is derived in a 
model independent way and in arbitrary spacetime dimensions $D$ by
solving the equations of motion of gravity coupled to a dilaton and
several different $n$-form field strengths. The intersection rules are
all compatible with supersymmetry, although derived without using it.
We then specialize to the branes occurring in type II string theories and
in M-theory. We show that the intersection rules are consistent with
the picture that open branes can have boundaries on some other branes.
In particular, all the D-branes of dimension $q$, with $1\leq q \leq6$,
can have boundaries on the solitonic 5-brane.

\end{abstract}

\newpage

\section{Introduction}
There has been recently considerable progress in the study of classical 
solutions of supergravities in 10 and 11 dimensions which are the
low-energy effective field theories of string theories and (the 
would-be) M-theory. These solutions play a key r\^ole for probing
the duality conjectures \cite{hulltown,democracy,witten} which appear
to relate between them all the string theories and M-theory.  
It is therefore important to gain better understanding of these
classical $p$-brane solutions.

In Type II string theories, there are two kinds of $p$-branes, those
charged under the NSNS fields and those carrying RR charge. The first ones
correspond to elementary states of string theory (for the 1-brane
\cite{dabholkar}) and to purely solitonic objects (for the 5-brane
\cite{callan_5brane}), while the second ones have been shown to
be described by D-branes \cite{polchinski,pol_lectures}. 
In the case of M-theory, since it is still at a conjectural level,
the 2- and the 
5-brane do not have any description involving elementary quantum objects
(however, the 5-brane has been 
conjectured to behave as a D-brane for open elementary membranes 
\cite{strom_open,town_mbranes,becker}). 

In their low-energy effective field theory description, the single 
$p$-brane solutions have been described in \cite{maeda,horostrom}. Although they
look quite similar, especially in the Einstein frame, they already
show a very different behaviour depending on their coupling to the dilaton.
M-branes (the $p$-branes of M-theory), due to the absence of a dilaton
in the theory, have a regular horizon\footnote{
Actually the maximally extended manifold of the 2-brane is much similar
to the one of an extreme Reissner-Nordstr\"om black hole, with a 
curvature singularity hidden by a horizon, while the 
5-brane manifold is completely regular \cite{gibhortow}.}, 
while $D=10$ $p$-branes all have
a naked curvature singularity at the location of the ``would-be''
horizon (one exception
is the self-dual RR 3-brane which does not couple to the dilaton of the
type IIB theory).

Intersections of $p$-branes of any kind are objects of 
growing interest, mainly because when enough branes intersect
a stabilization of the dilaton and of the moduli is achieved. This 
leads to a finite semiclassical entropy of the resulting black hole
after compactification has been carried over. In some cases, it has then
been possible to identify and count the microscopic states responsible
for that entropy, using D-brane technology,
in complete agreement with the semiclassical result
\cite{stromvafa,callanmalda,horstrom,maldastrom,johnkhuri}.

It is interesting in its own respect to derive a general rule stating
how $p$-branes can intersect. Heuristic arguments involving string
theory representation of the branes and duality have been given in
\cite{strom_open,town_mbranes,pol_lectures}. These arguments rely
heavily on the D-brane picture and on the requirement that the configuration
preserves some supersymmetries. In \cite{papatown}, the solutions 
of \cite{guven} are interpreted as intersecting M-branes of the same
kind, following the $p-2$ intersection rule for $p$-branes. Based on this
work and on other known solutions, in \cite{tseytlin_harm} Tseytlin
formulates the harmonic superposition rule for intersecting $p$-branes
(see also \cite{kastor,balalarsen}),
starting from 11 dimensions and deriving the intersection rules
from compactification and dualities. From the arguments of 
\cite{strom_open,town_mbranes,becker}, the 5- and the 2-brane in $D=11$
are taken to intersect on a string\footnote{
There is a different solution to $D=11$ supergravity involving a
2-brane inside a 5-brane \cite{izquierdo,costa,costapapa}. 
Since it preserves 1/2 of
the supersymmetries, it is clearly an intersection of a different kind
of the ones considered here.}. From their supersymmetry conditions and
using T-duality
the resulting intersecting $p$-brane bound states have been recently
classified in \cite{bergshoeff}, where M-branes and D-branes are
considered. A derivation of the intersection rules not based on
supersymmetry arguments has been given in \cite{tseytlin_noforce}, 
asking that $p$-brane probes in $p'$-brane backgrounds feel no force
and can thus create bound states with vanishing binding energy. Still,
the model for the $p$-brane probe is different if this one is an
M-brane, a D-brane or an NSNS-brane.

In this paper we find $p$-brane intersection rules purely from the
(bosonic) equations of motion of the low-energy theory\footnote{
While this was being completed, a paper \cite{arefeva} appeared
where a similar approach to this problem is considered.}. Moreover,
these rules are even model-independent. What we actually do is
to solve the equations of motion for a particular ansatz which has
as a consequence extremality and zero binding energy. Not only the
harmonic superposition rule is recovered, but also a constraint on
the way the different $p$-branes mutually intersect. This constraint 
depends on the coupling to the dilaton of the field strength under 
which they are charged. The dependence of the 
intersection rules on the branes we are considering is thus reduced
to this single characteristic. The ansatz we take, essentially
reducing all the different functions of transverse space to $\N$ 
independent ones, where
$\N$ is the number of intersecting branes, also implies
preservation of some of the supersymmetries.  

It is worth pointing out the intersection rule between the solitonic 
5-brane and the D-branes. If $q$ is the dimension of the D-brane, then
its intersection with the NSNS 5-brane has dimensionality $q-1$
(provided $1\leq q\leq 6$; however D-branes with $q\geq 7$ are for some
aspects pathological, at least from the classical point of view). It is
thus tempting to speculate that the solitonic 5-brane acts as a D-brane 
for the D-branes. This representation of the 
solitonic 5-brane is consistent with the dimensional reduction of the
$D=11$ 5-brane, which is seen as a D-brane for open membranes. Indeed, a
2-brane ending on a 5-brane can be related, after compactification on a
transverse direction, to a D2-brane ending on a solitonic 5-brane.
Also, the intersection rules for D-branes between themselves 
are compatible with the picture of
an open $q$-brane having boundaries on a $(q+2)$-brane 
\cite{strom_open}.

The paper is organized as follows: in Section 2 a general model inspired from
the bosonic sector of $D=10$ or 11 supergravity is presented, along 
with the successive ans\"{a}tze on the metric which allow us to find
a solution to the equations of motion. The condition for this solution
to be consistent yields the intersection rules.
In Section 3, we specialize to M-theory and string theories and rederive
the intersection rules for all the branes occurring in these
theories. In Section 4, we
speculate about the implications of a unified picture of all the
known branes, most notably for what concerns state counting in 
black hole entropy problems.

\section{General intersecting $p$-brane solution}

As a starting point, we take a very general action including gravity,
a dilaton and ${\cal M}$ field strengths of arbitrary form degree and
coupling to the dilaton. The action reads:
\be
I= \frac{1}{16\pi G_D} \int d^D x \sqrt{-g} 
\left( R - \frac{1}{2} (\partial \phi)^2 -
\sum_A \frac{1}{2 n_A !}e^{a_A \phi}F_{n_A}^2 \right), \qquad 
A=1\dots {\cal M} \label{action}
\ee
The metric is expressed in the Einstein frame.

Although we take the space-time to have a
generic dimension $D$, this action is most suitable for describing
the bosonic part of $D=10$ or $D=11$ supergravities. 
In fact, in \rref{action} we did not write the various Chern-Simons type 
terms which occur in these theories. Nevertheless, the solutions we will
present below (in section 3) are also consistent solutions of the full
equations of motion including Chern-Simons terms. 

To describe
lower dimensional supergravities, we should include several scalar
fields as in \cite{khviengia}, but since we are indeed mainly
interested in $D=10$ or 11 we prefer to keep only one scalar field
for simplicity. Also, we will nevertheless be able to consider these
lower dimensional cases by compactification.

In order to specialize to $D=10$, we will have to take $a_A=-1$
for the NSNS 3-form field strength and $a_A=\frac{1}{2}(5-n_A)$ for
the field strenghts coming from the RR sector. For $D=11$ we will
simply have to put $a_A=0$ for the 4-form.

The equations of motion (EOM) can be put in the following form:
\be
{R^\mu}_\nu=\frac{1}{2}\partial^\mu \phi \partial_\nu \phi+ \sum_A
\frac{1}{2n_A!}e^{a_A\phi} \left(n_A F^{\mu\rho_2\ldots\rho_{n_A}}
F_{\nu\rho_2\ldots\rho_{n_A}}- \frac{n_A-1}{D-2} \delta^\mu_\nu
F_{n_A}^2\right), \label{einstein}
\ee
\be
\Box \phi =\sum_A \frac{a_A}{2n_A!}e^{a_A\phi}F_{n_A}^2 \label{dilaton},
\ee
\be 
\partial_{\mu_1}\left(\sqrt{-g}e^{a_A\phi}F^{\mu_1\ldots\mu_{n_A}}
\right)=0. \label{maxwell}
\ee
The last set of equations has to be supplemented by the statement that
the $n_A$-form is the field strength of an $(n_A-1)$-form potential.
The same condition is obtained imposing the Bianchi identities (BI) to
the field strengths:
\be
\partial_{[\mu_1}F_{\mu_2\ldots\mu_{n_A+1}]}=0. \label{bianchi}
\ee

We now specialize to a particular form of the metric, which is a slight
generalization of the $p$-brane ansatz:
\be
ds^2=-B^2 dt^2+\sum_{i,j} C_{(i)}^2 \delta_{ij}dy_i dy_j +
G^2 \delta_{ab}dx^a dx^b, \label{metric}
\ee
where $y_i$ are compact coordinates with
$i,j=1\ldots p$, $a,b=1\ldots D-p-1$ and $B$, $C_{(i)}$ and $G$
depend only on the overall transverse coordinates $x^a$. Since we
will allow for multi-center solutions, we cannot postulate spherical
symmetry in the overall transverse space. However, we will eventually
recover spherical symmetry when all the branes are located at the
same point in transverse space.
Also, we take a
diagonal metric, thus excluding after compactification the presence
of KK momentum or KK monopoles.

The overall transverse space has dimension equal to $D-p-1$. We have to
recall at this stage that for the branes of dimension $q$, with $q$ less
than $p$, present in the solution, we have to consider a ``lattice'' of
such $q$-branes in the transverse compact directions and then to average
over them assuming that the compact directions are very small.

Considering now the $n$-form field strengths, we can generally make two
kinds of ans\"{a}tze. The electric ansatz is done asking that the BI are
trivially satisfied, while for the magnetic one we ask instead that the EOM for
the field strength are trivially satisfied. A correct electric
ansatz for a $q+2$ form corresponding to an electrically charged $q$-brane
is the following:
\be
F_{ti_1\ldots i_q a}=\epsilon_{i_1\ldots i_q} \partial_a E, \qquad
E=E(x^a). \label{electric}
\ee
One can easily check that it satisfies \rref{bianchi}.

For a magnetically charged $q$-brane, one needs an ansatz for a $D-q-2$
form. A good one which automatically satisfies the EOM \rref{maxwell} is:
\be
F^{i_{q+1}\ldots i_p a_1\ldots a_{D-p-2}}=\frac{1}{\sqrt{-g}}
e^{-a\phi}\epsilon^{i_{q+1}\ldots i_p}\epsilon^{a_1\ldots a_{D-p-1}}
\partial_{a_{D-p-1}}\tilde{E}, \qquad \tilde{E}=\tilde{E}(x^a). \label{magnetic}
\ee
Also the dilaton depends only on overall transverse space, $\phi=\phi(x^a)$.

Let us now discuss in some detail the next ans\"{a}tze that we will make
in order to solve the EOM \rref{einstein}--\rref{maxwell}.
First of all, consider how many independent functions we have at hand.
As far as the metric is concerned, one must have at least $\N$ different
functions in the set $(B, C_{(1)},\ldots  C_{(p)})$ 
in order to be able to make the distinction
between the $\N$ intersecting branes 
(from now on, we slightly change the notations; the index $A$ runs over 
all the branes, magnetic and electric).
These functions are supplemented by the metric component
relative to the overall transverse space $G$, the dilaton $\phi$ and the
$\N$ functions appearing in the field strengths, $E_A$ or $\tilde{E}_A$.
Our ultimate goal will be to reduce the number of independent functions
to $\N$, which, leading to the harmonic superposition rule, is interpreted
as the requirement of vanishing binding energy for the $p$-brane bound
state.
A first step towards this goal is to impose the following constraint:
\be
BC_{(1)}\ldots C_{(p)}G^{D-p-3}=1. \label{firstansatz}
\ee
This constraint effectively expresses the metric component $G$ as a function
of the others, and can be interpreted physically as enforcing 
extremality (it can be checked on a single $p$-brane solution \cite{horostrom}).
A consequence of this relation is that the ${R^a}_b$ components of
the Einstein equations \rref{einstein} are reduced to algebraic ones. We
will see that they indeed play a central r\^ole.

Because of \rref{firstansatz}, the EOM considerably simplify and become:
\be
\partial_a\partial_a \ln B=\frac{1}{2}\sum_A\frac{D-q_A-3}{D-2}
S_A (\partial_a E_A)^2, \label{tteom}
\ee
\be
\partial_a\partial_a \ln C_{(i)}=\frac{1}{2}\sum_A\frac{\delta^{(i)}_A}{D-2}
S_A (\partial_a E_A)^2, \label{iieom}
\ee
\begin{eqnarray} 
\lefteqn{\frac{\partial_aB \partial_b B}{B^2}+\sum_i 
\frac{\partial_aC_{(i)}\partial_bC_{(i)}}{C_{(i)}^2}+(D-p-3)
\frac{\partial_aG\partial_bG}{G^2}+\delta_{ab}\partial_c\partial_c\ln G=}
\nonumber \\
&=& \frac{1}{2}\sum_A S_A\partial_a E_A \partial_b E_A -\frac{1}{2}
\partial_a \phi\partial_b \phi -\delta_{ab}\frac{1}{2}\sum_A
\frac{q_A+1}{D-2}S_A(\partial_c E_A)^2, \label{abeom}
\end{eqnarray}
\be \partial_a\partial_a \phi=-\frac{1}{2}\sum_A 
\varepsilon_A a_A S_A (\partial_a E_A)^2,
\label{phieom}
\ee
\be
\partial_a(S_A \partial_a E_A)=0, \qquad 
S_A=(BC_{(i_1)}\ldots C_{(i_{q_A})})^{-2}e^{\varepsilon_A a_A \phi}. 
\label{maxeom}
\ee
We have $\delta^{(i)}_A=D-q_A-3$ if the direction $y_i$ is longitudinal to the
brane labelled by $A$ and $\delta^{(i)}_A=-(q_A+1)$ if it is transverse. 
$\varepsilon_A=(+)$ if the corresponding brane is electrically charged,
and $\varepsilon_A=(-)$ if it is magnetic, and in this case we dropped
the tilde from $E_A$.

The following step is to reduce the number of independent function to $\N$.
We make it in such a way that the only relevant equations will be the 
\rref{maxeom}, the other determining only algebraic coefficients.
The ansatz we take is the following:
\be
E_A=l_A H_A^{-1}, \qquad \qquad S_A=H_A^2, \label{secondansatz}
\ee
where $l_A$ is a constant to be determined later.
It follows that \rref{maxeom} directly reduce to:
\be
\partial_a \partial_a H_A =0, \label{harmonic}
\ee
thus characterizing the solution by $\N$ harmonic functions, each 
corresponding to the charge of a particular $p$-brane.
Actually, the most general solution of \rref{harmonic} is:
\be
H_A=1+\sum_k \frac{c_A Q_{A,k}}{|x^a-x^a_k|^{D-p-3}}, \label{multicenter}
\ee
i.e. a multicenter solution, which existence is a consequence of the
no-force condition \cite{tseytlin_noforce} between parallel
branes satisfying the
ans\"atze we took here.

The functions $B$, $C_{(i)}$ and $e^\phi$ are taken to be products
of the $H_A$, and the consistency of the equations \rref{tteom},
\rref{iieom}, \rref{abeom} and \rref{phieom} requires the second
relation of the ansatz \rref{secondansatz}.

If we take:
\be 
\ln B = - \sum_A \frac{D-q_A-3}{D-2} \alpha_A \ln H_A, \label{bcond}
\ee
\be
\ln C_{(i)}=- \sum_A \frac{\delta^{(i)}_A}{D-2} \alpha_A \ln H_A, \label{cond}
\ee
\be
\phi = \sum_A \varepsilon_A a_A \alpha_A \ln H_A, \label{phicond}
\ee
leading also to:
\be
\ln G = \sum_A \frac{q_A+1}{D-2} \alpha_A \ln H_A, \label{gcond}
\ee
the $p+2\geq \N +1$ equations \rref{tteom}, \rref{iieom} and \rref{phieom}
imply the $\N$ conditions:
\be 
\alpha_A \partial_a \partial_a \ln H_A +\frac{1}{2} l_A^2
(\partial_a \ln H_A)^2=0. \label{alphalcond}
\ee
By virtue of \rref{harmonic}, these conditions in turn imply:
\be
\alpha_A=\frac{1}{2} l_A^2. \label{alphalcond2}
\ee

The last set of equations \rref{abeom} becomes:
\begin{eqnarray}
\lefteqn{\sum_{A,B}\alpha_A \alpha_B \partial_a \ln H_A \partial_b
\ln H_B \left[\frac{(D-q_A-3)(D-q_B-3)}{(D-2)^2}+ \sum_i
\frac{\delta^{(i)}_A\delta^{(i)}_B}{(D-2)^2}+\right.} \nonumber \\
& & \mbox{} + \left.(D-p-3) \frac{(q_A+1)(q_B+1)}{(D-2)^2}+
\frac{1}{2} \varepsilon_A a_A \varepsilon_B a_B\right]
= \sum_A \alpha_A \partial_a \ln H_A \partial_b\ln H_B. \label{master}
\end{eqnarray}
For this set of equations to be satisfied for independent $H_A$, one has
two sets of algebraic conditions to satisfy: the first set contains
a condition for each brane and fixes the factor $\alpha_A$, and the
second set contains a condition for each pair of distinct $p$-branes, and
fixes their intersection rules.

If \rref{master} is rewritten as:
\be
\sum_{A,B}(M_{AB}\alpha_A-\delta_{AB})\alpha_B 
\partial_a \ln H_A \partial_b\ln H_B=0, \label{master2}
\ee
then the first set of conditions is given by $M_{AA}\alpha_A=1$, which 
yields:
\be
\alpha_A=\frac{D-2}{\Delta_A}, \label{alpha}
\ee
where:
\be
\Delta_A=(q_A+1)(D-q_A-3)+\frac{1}{2}a_A^2(D-2). \label{delta}
\ee

This completes the description of the solution, which is indeed a
superposition of single branes according to the ``harmonic superposition
rule'' formulated in \cite{tseytlin_harm}:
\[
B=\prod_A H_A^{-\frac{D-q_A-3}{\Delta_A}}, \qquad
C_{(i)}=\prod_A H_A^{-\frac{\delta^{(i)}_A}{\Delta_A}},
\]
\be
G=\prod_A H_A^{\frac{q_A+1}{\Delta_A}}, \qquad
e^\phi=\prod_A H_A^{\varepsilon_A a_A\frac{D-2}{\Delta_A}},
\label{solution}
\ee
\[
E_A=\sqrt{\frac{2(D-2)}{\Delta_A}}H_A^{-1}. 
\]

Let us now look at the second set of equations implied by \rref{master2},
$M_{AB}=0$ for $A\neq B$. Suppose that the two branes involved, characterized
by $q_1$ and $q_2$, intersect over $\bar{q}\leq q_1, q_2$ dimensions.
Define also $\bar{p}=q_1+q_2-\bar{q}\leq p$. Then we have:
\begin{eqnarray*}
M_{12}&=& \frac{1}{(D-2)^2}\{ (D-q_1-3)(D-q_2-3)+ \bar{q}(D-q_1-3)
(D-q_1-3)+ \\
& & \mbox{} - (q_1-\bar{q})(D-q_1-3)(q_2+1) - (q_2-\bar{q})(q_1+1)
(D-q_2-3)+ \\
& & \mbox{}+ (p-\bar{p})(q_1+1)(q_2+1) + (D-p-3)(q_1+1)(q_2+1)\}
+\frac{1}{2} \varepsilon_1 a_1 \varepsilon_2 a_2 \\
& =& (\bar{q}+1)-\frac{(q_1+1)(q_2+1)}{D-2}+\frac{1}{2}
\varepsilon_1 a_1 \varepsilon_2 a_2.
\end{eqnarray*}

We thus have an equation giving the number of dimensions on which two
branes intersect, depending on their own dimension and on their respective
coupling to the dilaton:
\be
\bar{q}+1=\frac{(q_A+1)(q_B+1)}{D-2}-\frac{1}{2}
\varepsilon_A a_A \varepsilon_B a_B. \label{intrule}
\ee

\section{Intersections in $D=11$ and $D=10$ supergravities}

We now specialize \rref{intrule} to cases of interest in M-theory and
string theory.

\subsection{Intersection of M-branes}

For $D=11$ supergravity, the 4-form field strength gives rise to electric
2-branes and magnetic 5-branes. Since there is no dilaton in this theory,
we simply take $a=0$ for all the branes we will consider.

The relation \rref{intrule} becomes:
\be
\bar{q}+1=\frac{(q_A+1)(q_B+1)}{9}. \label{mbranesrule}
\ee 
This rule gives the expected results, confirming the ones in
\cite{papatown,tseytlin_harm}: we have $\bar{q}=0$ for $2\cap 2$ (i.e. two
membranes intersect on a point), $\bar{q}=3$ for $5 \cap 5$ and $\bar{q}=1$
for $2\cap 5$. This last result is a support to the conjecture that open
membranes can end on the magnetic 5-brane. All these rules are valid
for each pair of branes in the configuration, so one can build along these
lines any bound state of more than two branes.

Let us notify that the derivation in
section 2 implicitly assumes that the overall transverse space is 
asymptotically flat. Configurations such that $D-p-3\leq 0$ are thus 
excluded. In the $D=11$ framework, we see that the configuration
of two 5-branes intersecting on a string \cite{kastor}, 
though allowed by straightforward
supersymmetry arguments, is automatically excluded.

\subsection{Intersection of D-branes}

In $D=10$ the field strengths arising from the RR sector of the superstring
couple to the dilaton in such a way that $\varepsilon a=\frac{1}{2}(3-q)$
both for electrically and magnetically charged $q$-branes. For two such
D-branes, \rref{intrule} can be recast in the following form:
\be
q_A+q_B-2\bar{q}=4, \label{dbranerule}
\ee
which was already used in \cite{balalarsen}. In its original derivation
\cite{pol_lectures}, which requires the intersection of D-branes to preserve
some supersymmetries, the r.h.s of \rref{dbranerule} is $0 \bmod 4$, thus
allowing for a larger number of intersections; however, $q_A+q_B-2\bar{q}=0$
is just the superposition of two similar branes, while if 
$q_A+q_B-2\bar{q}=8$, again the overall transverse space is not 
asymptotically flat.

To summarize, if we adopt the notation $q_A\cap q_B=\bar{q}$, we
have the following intersections: 
\begin{itemize}
\item in type IIA theory: $6\cap2=2$, $4\cap 4=2$, $4\cap2=1$, $4\cap0=0$
and $2\cap2=0$. Note for instance that the configuration $6\cap0$ is not
allowed and that $6\cap4$ gives 3 in spite of the fact that the 
transverse space is only 2 dimensional.
\item in type IIB theory: $5\cap3=2$, $5\cap1=1$, $3\cap3=1$, $3\cap1=0$.
Note here that $1\cap1$ gives $-1$, which could be reasonable in a 
Euclidean setting in which D-instantons are indeed present.
\end{itemize}
Note that from the results above we have the rule $(q+2)\cap q=q-1$ for 
D-branes (with $1\leq q \leq 3$) in agreement with \cite{strom_open}.

\subsection{Intersection of NSNS branes with other branes}

Let us first look at the intersections of the NSNS branes between themselves.
The NSNS 3-form field strength couples to the dilaton with $a=-1$, thus leading
to $1\cap1=-1$, $1\cap5=1$ and $5\cap5=3$ (even if in this case $D-p=3$)
for the NSNS 1- and 5-brane, in agreement with the $SL(2,Z)$ duality of type IIB
theory.

It is more instructive to check the intersection rules between NSNS branes and
D-branes. For the ``elementary'' string, we get:
\be 
1_{NS}\cap q_{RR}=0, \label{1branerule}
\ee
which states nothing else than the fact that open strings end on the D-branes.
For the solitonic 5-brane, we have \cite{tseytlin_noforce}:
\be
5_{NS}\cap q_{RR}=q-1, \label{5branerule}
\ee
leading us to speculate that the NSNS 5-brane can effectively act as the locus
on which the boundaries of open D-branes are constrained, i.e. it is a
D-brane for D-branes. This picture has to be confirmed by a calculation
which goes beyond the purely classical approach we are taking here.

Note that we did not take into account D-branes with $q> 6$ since from
the classical solution point of view they are rather pathological, yielding
automatically a transverse space of spatial dimension lower than 3. See
however \cite{bergshoeff} for a classification of intersecting branes
in $D-p=2$ and 3.

\section{Conclusion and discussion}

The aim of this paper was to uncover general intersection rules for the
$p$-branes which would go beyond their distinction between NSNS branes,
D-branes and even M-branes. Indeed, the charges carried by NSNS branes or
by D-branes appear to be related by U-duality\footnote{
To be complete, one should also add all the charges generated by the
KK reduction. We did not take them into account here for simplicity.}
\cite{hulltown}, while they are realized in a completely different way in string
theory. As a consequence, the microscopic state counting of the black hole
entropy is only possible for very particular configurations.
The conjecture that some branes could act as
D-branes for other branes, already formulated in \cite{strom_open},
might allow some steps ahead.

A picture in which the M-branes are treated in a way much similar to
the D-branes in string theory has also been used to perform some
counting of states in \cite{dijkgraaf,klebanov}. This is also consistent
with the relation between M-theory and the string theories.
Here we have shown that, from the point of view of the classical 
solutions, in which U-duality is a true symmetry of the equations of motion,
if we interpret the result \rref{1branerule} as a hint for the existence
of D-branes for strings, then we have also to suppose the existence of
D-branes for higher branes. The consistency of the open membrane picture
in M-theory with compactification also requires this.

We can calculate the mass of the solutions given in \rref{solution} by using
the formula for the ADM mass:
\be
M=-\frac{L^p \Omega_{D-p-2}}{8\pi G_D}r^{D-p-2}[\sum_i \partial_r C_{(i)} +
(D-p-2) \partial_r G]|_{r\rightarrow \infty}. \label{adm}
\ee
Using \rref{solution} and the fact that $H_A \rightarrow 1$ when
$r\rightarrow \infty$, one obtains:
\be
M=\sum_A M_A, \label{zerobinding}
\ee
where $M_A=Q_A$ in suitable units
for each constituent $p$-brane. The result \rref{zerobinding}
comes essentially from the second ansatz \rref{secondansatz}, assuring
zero binding energy, while the extremality condition for each brane
is due to the first ansatz \rref{firstansatz}.

One can check that the configurations \rref{solution} are supersymmetric.
The exact amount of preserved supersymmetry depends on each particular 
configuration and is at least equal to $1/2^{\N}$.  
The interest of the approach followed here is its independence with 
respect to the model and the space-time dimension.
On the contrary, the approach based on supersymmetry depends heavily 
on the particular supergravity model considered.

Since the harmonic functions have the form $H_A=1+\frac{c_A Q_A}{r^{D-p-3}}$,
we can calculate the behaviour of the area and of the dilaton at the
`horizon'. For the area:
\begin{eqnarray*}
A_{D-p-2}&=&\Omega_{D-p-2} r^{D-p-2}C_{(1)}\ldots C_{(p)}G^{D-p-2}\\
&=&\Omega_{D-p-2}r^{D-p-2}B^{-1} G\\
&=&\Omega_{D-p-2}r^{D-p-2}\prod_A H_A^{
\frac{D-2}{\Delta_A}}|_{r=0}. 
\end{eqnarray*}
For $D=10$ or $D=11$ supergravities, we always have 
$\frac{D-2}{\Delta_A}=\frac{1}{2}$ and thus we need 4 charges to have
a 4 dimensional black hole with non-zero entropy and 3 charges for a
5 dimensional one. It is worth pointing out that the formula above
states that these two cases are really the only ones which allow for
a non-zero extremal entropy, at least in the framework of configurations
in $D=11$ or 10 supergravity without internal momenta.

For the dilaton to be constant at the horizon, we see in \rref{solution}
that one needs to check for each configuration that
$\sum_A \varepsilon_A a_A \frac{D-2}{\Delta_A}=0$.
One indeed always finds
that when the area of the horizon is finite, the dilaton
has a fixed value at the horizon. 

It is puzzling that the entropy has a microscopic explanation only for
particular configurations involving D-branes and momentum.
Here, we considered extremal black hole configurations
built up exclusively by branes, i.e. without momenta in the internal 
directions. One such configuration which yields a 5 dimensional black hole
is the highly symmetric $2\cap2\cap2$ intersection in M-theory. After 
compactification and a chain of T-dualities, this solution is related to
the ``$1_R\cap5_R$ + momentum'' configuration of type IIB theory, which
was used to perform a counting of microstates in \cite{callanmalda,horstrom}.
The counting critically uses the presence of a momentum in the bound 
state.
It would be nice to be able to perform a counting of microstates
for the configuration when it is expressed in its most simple form.

\subsection*{Acknowledgments}
We would like to thank all the
participants in the work group on the recent developpements in string
and black hole physics held in the Laboratoire de Physique Th\'eorique et
Hautes Energies in Paris, and especially I.~Antoniadis, C.~Bachas, 
T.~Damour and P.~Windey for stimulating discussions.


\begin{thebibliography}{99}

\bibitem{hulltown} C.~M.~Hull and P.~K.~Townsend, ``Unity of Superstring 
Dualities", Nucl. Phys.  {\bf B438} (1995) 109; hep-th/9410167.

\bibitem{democracy}  P.~K.~Townsend, ``P-Brane Democracy", 
proceedings of the March 95 PASCOS/John Hopkins Conference; 
hep-th/9507048. 

\bibitem{witten} E.~Witten, ``String Theory Dynamics in 
Various Dimensions", Nucl. Phys.  {\bf B443} (1995) 85;  
hep-th/9503124.

\bibitem{dabholkar} A.~Dabholkar, G.~W.~Gibbons, J.~A.~Harvey and 
F.~Ruiz-Ruiz, ``Superstrings and Solitons'', 
Nucl. Phys.  {\bf B340} (1990) 33. 



\bibitem{callan_5brane} C.~G.~Callan, J.~A.~Harvey and A.~Strominger, 
``Worldsheet Approach to Heterotic Instantons and Solitons', 
Nucl. Phys.  {\bf B359} (1991) 611.

\bibitem{polchinski} J.~Polchinski, 
``Dirichlet-Branes and Ramond-Ramond Charges'', 
Phys. Rev. Lett. {\bf 75} (1996) 4724; 
 hep-th/9510017.

\bibitem{pol_lectures} J.~Polchinski, S.~Chaudhuri and C.~V.~Johnson,
``Notes on D-Branes", hep-th/9602052;
J.~Polchinski, ``TASI Lectures on D-Branes'', hep-th/9611050. 


\bibitem{strom_open} A.~Strominger, ``Open P-Branes", 
 Phys. Lett. {\bf B383} (1996) 44; hep-th/9512059.

\bibitem{town_mbranes} P.~K.~Townsend, ``D-branes from M-branes", 
 Phys. Lett. {\bf B373} (1996) 68; hep-th/9512062.

\bibitem{becker} K.~Becker and M.~Becker, ``Boundaries in  M-Theory",  
 Nucl. Phys. {\bf B472} (1996) 221; hep-th/9602071.

\bibitem{maeda} G.~W.~Gibbons and K.~Maeda, 
``Black Holes and Membranes in Higher-Dimensional Theories with 
Dilaton Fields'', 
 Nucl. Phys. {\bf B298} (1988) 741.

\bibitem{horostrom} G.~T.~Horowitz and A.~Strominger, 
``Black Strings and p-Branes'', 
 Nucl. Phys. {\bf B360} (1991) 197.

\bibitem{gibhortow} G.~W.~Gibbons, G.~T.~Horowitz and P.~K.Townsend, 
``Higher-dimensional resolution of dilatonic black-hole singularities'',
 Class. Quantum Grav. {\bf 12} (1995) 297; hep-th/9410073.

\bibitem{stromvafa} A.~Strominger and  C.~Vafa, 
``Microscopic Origin of the Bekenstein-Hawking Entropy'', 
 Phys. Lett. {\bf B379} (1996) 99; hep-th/9601029.

\bibitem{callanmalda} C.~G.~Callan and J.~M.~Maldacena, 
``D-brane Approach to Black Hole Quantum Mechanics", 
 Nucl. Phys. {\bf B472} (1996) 591; hep-th/9602043.

\bibitem{horstrom} G.~Horowitz and A.~Strominger, 
``Counting States of Near-Extremal Black Holes", 
Phys. Rev. Lett. {\bf 77} (1996) 2368; hep-th/9602051.


\bibitem{maldastrom} J.~M.~Maldacena and A.~Strominger, 
``Statistical Entropy of Four-Dimensional Extremal Black Holes", 
Phys. Rev. Lett. {\bf 77} (1996) 428; hep-th/9603060.

\bibitem{johnkhuri} C.~V.~Johnson,  R~.R.~Khuri and R.~C.~Myers, 
``Entropy of 4D Extremal Black Holes", 
 Phys. Lett. {\bf B378} (1996) 78; hep-th/9603061.



\bibitem{papatown} G.~Papadopoulos and P.~K.~Townsend, 
`` Intersecting M-branes", Phys. Lett. {\bf B382} (1996) 65; 
hep-th/9603087.

\bibitem{guven} R.~G\"uven, ``Black p-Brane Solutions of D=11 
Supergravity Theory'', Phys. Lett. {\bf B276} (1992) 49.


\bibitem{tseytlin_harm} A.~A.~Tseytlin, 
``Harmonic superpositions of M-branes'', 
Nucl. Phys. {\bf B475} (1996) 149;  
hep-th/9604035.

\bibitem{kastor} J.~P.~Gauntlett, D.~A.~Kastor and J.~Traschen, 
``Overlapping Branes in M-Theory'',  Nucl. Phys.
  {\bf B478} (1996) 544; hep-th/9604179.

\bibitem{balalarsen} V.~Balasubramanian and F.~Larsen,
``On D-Branes and Black Holes in Four Dimensions'',
Nucl. Phys. {\bf B478} (1996) 199; hep-th/9604189.



\bibitem{izquierdo} J.~M.~Izquierdo, N.~D.~Lambert,
 G.~Papadopoulos and P.~K.~Townsend, `` Dyonic Membranes", Nucl. Phys. 
  {\bf B460} (1996) 560; hep-th/9508177.

\bibitem{costa} M.~S.~Costa, ``Composite M-branes", hep-th/9609181.

\bibitem{costapapa}  M.~S.~Costa and G.~Papadopoulos, 
``Superstring dualities and p-brane bound states", hep-th/9612204.

\bibitem{bergshoeff}  E.~Bergshoeff, M.~de Roo, E.~Eyras, B.~Janssen and 
J.~P.~van der Schaar, ``Multiple Intersections of D-branes and M-branes"
, hep-th/9612095.


\bibitem{tseytlin_noforce} A.~A.~Tseytlin, `` No-force condition and BPS 
combinations of p-branes in 11 and 10 dimensions", hep-th/9609212.

\bibitem{arefeva} I.~Y.~Aref'eva and O.~A.~Rytchkov, ``Incidence 
Matrix Description of Intersecting p-brane Solutions", hep-th/9612236.

\bibitem{khviengia} N.~Khviengia, Z.~Khviengia, H.~L\"u and C.~N.~Pope, 
``Intersecting M-branes and bound states'', 
 Phys. Lett. {\bf B388} (1996) 21; 
hep-th/9605077.


\bibitem{dijkgraaf} R.~Dijkgraaf, E.~Verlinde and H.~Verlinde, 
``BPS Spectrum of the Five-Brane and Black Hole Entropy'', 
hep-th/9603126; ``BPS Quantization of the Five-Brane", hep-th/9604055.

\bibitem{klebanov} I.~R.~Klebanov and  A.~A.~Tseytlin, 
``Intersecting M-branes as Four-Dimensional Black Holes'', 
Nucl. Phys. {\bf B475} (1996) 179; hep-th/9604166.


\end{thebibliography}
\end{document}